\documentclass{article}[12pt]
\usepackage{enumerate}
\usepackage{amsmath}
\usepackage{dcolumn}
\usepackage{graphicx}
\usepackage{bm}
\begin{document}
\title{Black Hole evaporation in semi-classical approach}

\author{Shintaro Sawayama\footnote{email:sawayama0410@gmail.com} \\
Sawayama Cram School of Physics \\
Atsuhara 328, Fuji-shi, Shizuoka-ken, Japan, 419-0201 }
\maketitle
\begin{abstract}
As well as known, the black hole evaporation problem is famous problem.
Because the S.W.Hawking found the black holes emit light at the future null infinity as a thermal radiation \cite{H},
we think that the black holes may be vanish.
However, to prove this problem, we should solve field equation, i.e. forth order partial differential equations \cite{Ford}\cite{BD}.
However, we can find a method to solve this equation, and we could prove that the black holes finally vanish.
To solve this problem we use dynamical horizons equation in the Vaidya spacetime.
\end{abstract}
\section{Introduction}\label{sec1}

\quad We can show the black holes finally vanish by the analytical method and by the dynamical horizon equations.
As we proved paper at 2006 in \cite{Sa}, the black holes finally vanish, by solving the dynamical horizon equations, and present paper mainly based on the paper.
Note that the dynamical horizon equations and the Einstein equations are same, and to solve dynamical horizon equations is same as to solve field equation.
However, in the previous paper, we use the approximations that we use the negative energy \cite{C} near the event horizon not for the dynamical horizons.
So in this paper we do not use the approximations.
Our paper explain the result of previous paper and we show black hole evaporation problem without any approximations.

The black hole evaporation problem starts from S.W.Hawking \cite{H} at 1975.
By his study we could find the black holes emit lights as the thermal radiation.
And we can think that the black holes may be vanished by this effect i.e. Hawking effect.
However, the proof of this problem seems to be hard, because we should solve field equations,
and the field equations are forth order partial differential equations with non-linear term.
We can not find the mathematical methods to solve such equations.
However, by the study of the dynamical horizons \cite{AK1}\cite{AK2}, we can solve this problem.

We can explain shortly the previous works of the black hole evaporation.
The orthodox work may be Page's result \cite{Page} as $\dot{M}\propto -M^{-2}$, it comes from the dimensional analysis.
The string approach is \cite{HK}\cite{GR}\cite{TT},
or semiclassical theory typically using apparent horizon \cite{PA}.
Hiscok studied spherical model of the black hole evaporation using the Vaidya metric, which we also use in present work, 
to solve the black hole evaporation problem.
However, he simply set a model not taking account of the field equation. 
Hajicek's work\cite{Haji} treated the black hole mass more generally than our present case. 
However, he did not use the field equation either. 
One of the more recent studies is Sorkin and Piran's work \cite{SP} on charged black holes.
And neutral case has been done by Hamade and Stewart\cite{HS}.
Their conclusion is that black hole mass decreases or increases depending on initial condition.
They used a model of the double null coordinates, and obtained a numerical result.
But they did not consider the Hawking effect directly but they used massless scalar field as a matter.
Brevik and Halnes calculated primordial black hole evaporation\cite{BH}.
Very recently Hayward studied black hole evaporation and formation using the Vaidya metric \cite{Hay}.  
It seems no analytical equation has been proposed for the black hole mass with the Hawking effect taken into account.

In a section \ref{sec2} we explain our previous result \cite{Sa} by short sentences.
And in a section \ref{sec3}, we explain the calculations of the dynamical horizons equation without any approximations.
Finally in a section \ref{sec4}, we discuss and conclude our results.

\section{Previous results by short sentences}\label{sec2}
\quad We can show our previous result in three subsections.
In the subsection \ref{sec2-1}, we explain the dynamical horizons equation.
In the subsection \ref{sec2-2}, we explain the Vaidya spacetime.
And in the subsection \ref{sec2-3}, we show our previous result by calculating the dynamical horizons equation.
The subsection \ref{2-2} and \ref{sec2-3} is the review of the previous paper and mainly same as \ref{Sa}.
The reason why we use the dynamical horizons equation is it only needs information of the horizon surface, and reason why we use the Vaidya spacetime is that the dynamics of the horizon is free by the fluent matter.
\subsection{Dynamical horizons}\label{sec2-1}
\quad The study of the Dynamical horizon is mainly depend on Ashtekar and Krishnan's works \cite{AK1}\cite{AK2}, 
and we omit the main calculations to derive the dynamical horizons equation.
The definition of the dynamical horizons are as follows. \\ \\

{\it Definition}. A smooth, three-dimensional, spacelike submanifold $H$ in a space-time
is said to be a {\it dynamical horizon} if it is foliated by preferred family of 2-spheres such that, 
on each leaf S, the expansion $\Theta _{(l)}$ of a null normal $l^a$ vanishes and the expansion $\Theta_{(n)}$
 of the other null normal $n^a$ is strictly negative. \\ \\

This definition is similar to the apparent horizons.
For simplicity, the definition of the black hole is that the black hole does not emit even light.
Although the above definition was given by Ashtekar and Krishnan, the definition define only spacelike horizons.
So we can modify the above definition, as follows. \\ \\

{\it Definition (modified version)}. A smooth, three-dimensional, spacelike or {\bf timelike} submanifold $H$ in a space-time
is said to be a {\it dynamical horizon} if it is foliated by preferred family of 2-spheres such that, 
on each leaf S, the expansion $\Theta _{(l)}$ of a null normal $l^a$ vanishes and the expansion $\Theta_{(n)}$
 of the other null normal $n^a$ is strictly negative. \\ \\

The timelike dynamical horizon is also called timelike membranes.

There is the dynamical horizons equation.
From the Einstein equation on the dynamical horizons, we can use 3+1 decomposition and then 2+1 decomposition, and finally using the Gauss-Bonet theorem, we obtain
\begin{eqnarray}
(\frac{R_2}{2G}-\frac{R_1}{2G})=\int _{\Delta H}\bar{T}_{ab}\hat{\tau}^a\xi ^b_{(R)}d^3V\nonumber \\
+\frac{1}{16\pi G}\int _{\Delta H}(|\sigma |^2+2|\zeta |^2)d^3V. \label{eq1}
\end{eqnarray}
Here, $R_1$ is the radios of the dynamical horizons before the energy flows in the horizons and $R_2$ is the radius of the dynamical horizons after the energy flows in the horizons.
And $G$ is the gravitational constant, and integration is on the dynamical horizon $\Delta H$, and $\bar{T}_{ab}$ means stress-energy tensor with cosmological constant, and 
the $\hat{\tau}^a$ is the normal timelike vector diagonal to horizons and $\xi ^b_{(R)}$ is the $N_R=|\partial R|$ where $R$ is the black hole radius.
And the second term of the integrand is the form of the Bondi energy.
However, the second term become vanish because we only treat spherically symmetric spacetime i.e. the Vaidya spacetime.
Then the equation (\ref{eq1}) becomes as,
\begin{eqnarray}
(\frac{R_2}{2G}-\frac{R_1}{2G})=\int _{\Delta H}\bar{T}_{ab}\hat{\tau}^a\xi ^b_{(R)}d^3V. \label{eq2}
\end{eqnarray}
In this paper we use this simple equation.
We can easily calculate the timelike dynamical horizon equation only replacing $\hat{r}^a$ and $\hat{\tau}^a$. 
And the way of 3+1 and 2+1 decomposition is replaced.
\subsection{Vaidya spacetime}\label{sec2-2}
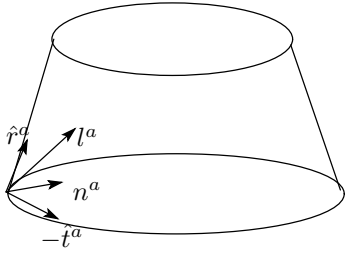
\begin{figure}
\unitlength 0.1in
\begin{picture}( 17.7000, 12.1000)(  7.9000,-18.5000)
%
\special{pn 8}%
\special{ar 1660 830 630 190  0.0000000 6.2831853}%
%
\special{pn 8}%
\special{ar 1680 1640 880 210  0.0000000 6.2831853}%
%
\special{pn 8}%
\special{pa 1040 830}%
\special{pa 800 1640}%
\special{fp}%
%
\special{pn 8}%
\special{pa 2280 860}%
\special{pa 2540 1620}%
\special{fp}%
%
\special{pn 8}%
\special{pa 790 1630}%
\special{pa 900 1360}%
\special{fp}%
\special{sh 1}%
\special{pa 900 1360}%
\special{pa 856 1414}%
\special{pa 880 1410}%
\special{pa 894 1430}%
\special{pa 900 1360}%
\special{fp}%
%
\special{pn 8}%
\special{pa 790 1630}%
\special{pa 1060 1770}%
\special{fp}%
\special{sh 1}%
\special{pa 1060 1770}%
\special{pa 1010 1722}%
\special{pa 1014 1746}%
\special{pa 992 1758}%
\special{pa 1060 1770}%
\special{fp}%
%
\special{pn 8}%
\special{pa 800 1620}%
\special{pa 1150 1300}%
\special{fp}%
\special{sh 1}%
\special{pa 1150 1300}%
\special{pa 1088 1330}%
\special{pa 1112 1336}%
\special{pa 1114 1360}%
\special{pa 1150 1300}%
\special{fp}%
%
\special{pn 8}%
\special{pa 790 1630}%
\special{pa 1080 1580}%
\special{fp}%
\special{sh 1}%
\special{pa 1080 1580}%
\special{pa 1012 1572}%
\special{pa 1028 1590}%
\special{pa 1018 1612}%
\special{pa 1080 1580}%
\special{fp}%
\put(11.6000,-13.9000){\makebox(0,0)[lb]{$l^a$}}%
\put(7.9000,-13.8000){\makebox(0,0)[lb]{$\hat{r}^a$}}%
\put(9.7000,-19.3000){\makebox(0,0)[lb]{$-\hat{t}^a$}}%
\put(11.4000,-16.7000){\makebox(0,0)[lb]{$n^a$}}%
\end{picture}
\caption{For the case that the dynamical horizon decreases, we should choose $l^a=-\hat{t}^a+\hat{r}^a$ 
so that $l^a$ points into the dynamical horizon.}\label{fig5}
\end{figure}

\quad The Vaidya metric is of the form
\begin{eqnarray}
ds^2=-Fdv^2+2Gdvdr+r^2d\Omega ^2 ,
\end{eqnarray}
where $F$ and $G$ are functions of $v$ and $r$,
and $v^a$ is null vector and $r$ is the area radius, and $M$ is the mass defined by $M=\frac{r}{2}(1-\frac{F}{G^2})$, 
a function of $v$ and $r$. 
This metric is spherically symmetric.
By substituting the Vaidya metric (2) into the Einstein equation so that we can identify the energy-momentum tensor $T_{ab}$ as  
\begin{eqnarray}
8\pi T_{vv}&:=&\frac{2}{r^2}(FM_{,r}+GM_{,v}) \\
8\pi T_{vr}&:=&-\frac{2G}{r^2}M_{,r} \\
8\pi T_{rr}&:=&\frac{2G_{,r}}{rG}.
\end{eqnarray}
We do not need to check that the solution of the dynamical horizon equation satisfies the Einstein equation.
Because we would like to consider the Schwarzschild like metric, we set $v=t+r^*$, 
where $r^*$ is tortoise coordinate with dynamics
\begin{eqnarray}
r^*=r+2M(v)\ln \bigg( \frac{r}{2M(v)}-1\bigg) .
\end{eqnarray}
For later convenience, we write,
\begin{eqnarray}
a=\frac{d r}{d r^*}.
\end{eqnarray}
There are two null vectors,
\begin{eqnarray}
l^a=\begin{pmatrix}
l^t \\
l^{r*}\\
l^{\theta} \\
l^{\phi}
\end{pmatrix}=
\begin{pmatrix}
-a^{-1}\\
a^{-1}\\
0\\
0
\end{pmatrix},
\end{eqnarray}
corresponding to the null vector $v^a$,
and the other is
\begin{eqnarray}
n^a=\begin{pmatrix}
n^t\\
n^{r*}\\
n^{\theta}\\
n^{\phi}
\end{pmatrix}=
\begin{pmatrix}
-a^{-1}\\
-\frac{F}{F-2Ga}a^{-1}\\
0\\
0
\end{pmatrix}.
\end{eqnarray}
Here we multiply $a^{-1}$ so that $l^a=v^a$. 
This choice of the null vector $l^a$ is explained in figure \ref{fig5}.
From now on we put,
\begin{eqnarray}
F&=&\bigg( 1-\frac{2M(v)}{r}\bigg) \\
G&=&1,
\end{eqnarray}
in a similar form to the Schwarzschild metric, 
assuming that $M(v)$ is a function of $v$ only. 
For a constant $M$, the metric coincides with the Schwarzschild metric.
We calculate the expansions $\Theta _{(l)}$ and $\Theta _{(n)}$ of the two null vectors $l^a,n^a$,  
because the definition of the dynamical horizon requires one of the null expansions to be zero and the other to be minus. 
The result is,
\begin{eqnarray}
\Theta _{(l)}&=&\frac{1}{r}(2F-a)\\
\Theta _{(n)}&=&\frac{1}{ar}\bigg( \frac{-2F^2+aF-2a^2}{-F+2a}\bigg) .
\end{eqnarray}
From $\Theta _{(l)}=0$ we get,
\begin{eqnarray}
2F-a=0.
\end{eqnarray}
we can check that the other null expansion $\Theta_{(n)} $is strictly negative. 
Therefore in this case, we can apply the dynamical horizon equation. 
In the usual Schwarzschild metric with dynamics, both expansions become zero. 
This is the one of the reasons why we choose the Vaidya metric. 
By inserting equation (7) to equation (14), we obtain
\begin{eqnarray}
a=F\bigg( 1-2M_{,v}\ln \bigg(\frac{r}{2M}-1\bigg) + \frac{r}{M(r/2M-1)}M_{,v}\bigg) .
\end{eqnarray}
Note that $a$ is proportional to $F$. 
Now we solve $\Theta _{(l)}=0$, to determine the dynamical horizon radius as
\begin{eqnarray}
2F-a=2F
-F\bigg( 1-2M_{,v}\ln \bigg(\frac{r}{2M}-1\bigg) 
+ \frac{r}{M(r/2M-1)}M_{,v}\bigg) =0.
\end{eqnarray}
From this equation we obtain,
\begin{eqnarray}
1+\bigg( -2M_{,v}\ln \bigg(\frac{r_D}{2M}-1\bigg) 
+ \frac{r_D}{M(r_D/2M-1)}M_{,v}\bigg) =0.
\end{eqnarray}
Here $F=0$ is also the solution of the dynamical horizon.
The dynamical horizon radius $r_D$ is given by solving (17) as
\begin{eqnarray}
r_D=2M+2Me^{-v/2M}. \label{f43}
\end{eqnarray}
Note that this dynamical horizon radius is outside the $r=2M$, that is othor solution.
\subsection{Black hole evaporation problem with approximation}\label{sec2-3}
\begin{figure}
\includegraphics[width=50mm]{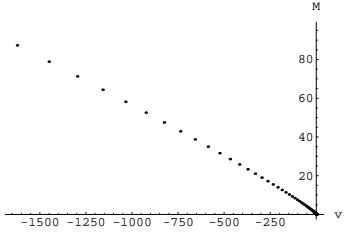}
\caption{Numerical calculation of the black hole mass $M$ as a function of v from the equation (\ref{f4})}\label{plot}
\end{figure}

At first, we should derive the energy-momentum tensor $T_{\hat{t}l}$ 
for the integration of the dynamical horizon equation. 
For this end we derive it from the given Vaidya matter. 
For $G=1, \ F=1-\frac{2M(v)}{r}$, the non-vanishing components of the energy-momentum tensor becomes 
\begin{eqnarray}
T_{vv}&=&\frac{1}{4\pi r^2}(FM_{,r}+M_{,v}) \label{f45}\\
T_{lr^*}&=&-\frac{1}{4\pi r^2}M_{,r}a \label{f46}\\
T_{r^*r^*}&=&0.
\end{eqnarray}
Here we have made the coordinate transformation from $r$ to $r^*$.
Writing $T_{tl}$ in terms of $T_{vv}$ and $T_{vr^*}$ given by (\ref{f45})(\ref{f46}) with $l^a=v^a$, we see
\begin{eqnarray}
T_{tl}&=&-T_{vv}+T_{vr^*} \nonumber \\
&=&\frac{1}{4\pi r^2}(-FM_{,r}-M_{,v}-aM_{,r}) \nonumber \\
&=&-\frac{1}{4\pi r^2}\frac{5}{2}M_{,v}.
\end{eqnarray}
With $\hat{t}^a$ being the unit vector in the direction of $t^a$, we obtain
\begin{eqnarray}
T_{\hat{t}l}=-\frac{1}{4\pi r^2}\frac{5}{2}M_{,v}F^{-1}.
\end{eqnarray}
For the dynamical horizon integration (\ref{eq2}), 
we get
\begin{eqnarray}
\int _{r_1}^{r_2}4\pi r_D^2T_{\hat{t}l}dr_D=\frac{5}{2}\int _{M_1}^{M_2}(1+e^{-v/2M})dM, \label{f52}
\end{eqnarray}
where we have used
\begin{eqnarray}
F=\frac{e^{-v/2M}}{1+e^{-v/2M}},
\end{eqnarray}
and the fact
\begin{eqnarray}
\frac{dM}{dv}=-e^{-v/2M}\bigg( 2(1+e^{-v/2M})+\frac{v}{M}e^{-v/2M}\bigg) ^{-1}, \label{eq7}
\end{eqnarray}
changing the integration variable from $r_D$ to $M$. 
In the above calculation, we treat $M_{,v}$ and $F^{-1}$ with $r_D$ fixed, 
because these functions are used only in the integration.
Inserting equation (\ref{f52}) to the dynamical horizon equation (\ref{eq2}), we obtain
\begin{eqnarray}
\frac{1}{2}(2M+2Me^{-v/2M})\bigg| _{M_1}^{M_2} 
=\int _{M_1}^{M_2}\frac{5}{2}(1+e^{-v/2M})dM .
\end{eqnarray}
Taking the limit $M_2\to M_1=M$, we obtain
\begin{eqnarray}
-\frac{3}{2}(1+e^{-v/2M})+\frac{v}{2M}e^{-v/2M}=0 .
\end{eqnarray}
This equation is the dynamical horizon equation in the case that only the Vaidya matter is present. 
There is no solution of this equation, except the trivial one ($F=0$ or $r=2M$), so
\begin{eqnarray}
r_D=2M(v).
\end{eqnarray}
Here $M(v)$ is the arbitrary function only of the $v$, which represent the Vaidya black hole spacetime. \\
\ \ \ Next, we take into account the Hawking radiation. 
To solve this problem, we use two ideas that is to use the dynamical horizon equation, and 
to use the Vaidya metric. 
The reason to use the dynamical horizon equation comes from the fact that we need only information of matter near horizon,  
without solving the full Einstein equation with back reaction being the fourth order differential equations, 
for a massless scalar field.
For the matter on the dynamical horizon, we use the result of Candelas \cite{C}, 
which assumes that 
spacetime is almost static and is valid near the horizon, $r\sim 2M$.
\begin{eqnarray}
T_{tl}&=&-T_{tt}\nonumber \\
&=&\frac{1}{2\pi ^2(1-2M/r)}\int _0^{\infty}\frac{d\omega \omega ^3}{e^{8\pi M\omega}-1} \nonumber \\
&=&\frac{1}{2cM^4\pi ^2(1-2M/r)},
\end{eqnarray}
where we have used a well known result, 
\begin{eqnarray}
\int _0^{\infty}\frac{d\omega \omega ^3}{e^{a\omega}-1}=\frac{\pi ^4}{15a^4},
\end{eqnarray}
and where $c=61440$.
This matter energy is negative near the event horizon. 
In the dynamical horizon equation, if black hole absorbs negative energy, 
black hole radius decreases. 
This is one of the motivations to use the negative energy tensor. 
Next we replace length of $t$ to unit length, because in the dynamical horizon equation $\hat{t}$ is used, so
\begin{eqnarray}
\hat{t}^{0}=F^{-1/2},\ \ l^{0}=F^{-1/2},
\end{eqnarray}
and therefore, the energy tensor becomes
\begin{eqnarray}
T_{\hat{t}l}=\frac{1}{2M^4c\pi ^2(1-2M/r)^2}.
\end{eqnarray}
Calculating the integration on the right hand side of (\ref{eq2}) for this matter,
\begin{eqnarray}
b\int \frac{r_D^2}{M^4(1-2M/r_D)^2}dr_D\nonumber \\
=b\int \frac{4M^2(1+e^{-v/2M})^4}{M^4e^{-v/M}}\frac{dr_D}{dM}dM \nonumber \\
=b\int _{R_1}^{R_2}\frac{4(1+e^{-v/2M})^4}{M^2}e^{-v/M} \bigg( 2(1+e^{-v/2M})+\frac{v}{M}e^{-v/2M}\bigg) dM.
\end{eqnarray}
Here we insert the expression for $r_D$ (\ref{f43})in the first line, and the expression for $dr_D/dM=2(1+e^{-v/2M})+\frac{v}{M}e^{-v/2M}$ is used. 
Here $b$ is a constant calculated in \cite{C}
\begin{eqnarray}
b=\frac{1}{30720\pi}.
\end{eqnarray}
If we also take account of the contribution of the Vaidya matter, 
and inserting this into the integration to the dynamical horizon equation (\ref{eq2}), we obtain
\begin{eqnarray}
\frac{1}{2}( 2M+2Me^{-v/2M})\bigg| _{M_1}^{M_2}\nonumber \\
=b\int _{M_1}^{M_2}\frac{2^2(1+e^{-v/2M})^4}{M^2}e^{-v/M} \bigg( 2(1+e^{-v/2M})+\frac{v}{M}e^{-v/2M}\bigg) dM \nonumber \\
+\int _{M_1}^{M_2}\frac{5}{2}(1+e^{-v/2M})dM.
\end{eqnarray}
Taking the limit $M_2\to M_1=M$, we finally get
\begin{eqnarray}
-\frac{3}{2}(1+e^{-v/2M})+\frac{v}{2M}e^{-v/2M}\nonumber \\ 
=b\frac{2^2(1+e^{-v/2M})^4}{M^2} e^{v/M}\bigg( 2(1+e^{-v/2M})+\frac{v}{M}e^{-v/2M}\bigg) , 
\end{eqnarray}
or
\begin{eqnarray}
M^2=\frac{8b(1+e^{-v/2M})^4e^{v/M}}{-\frac{3}{2}(1+e^{-v/2M})+\frac{v}{2M}e^{-v/2M}} \bigg( (1+e^{-v/2M})+\frac{v}{2M}e^{-v/2M}\bigg) .\label{f4}
\end{eqnarray}
This is the main result of the previous work that describes 
how the mass of black hole decreases. 
This equation is the transcendental equation, so usually it cannot be solved analytically. 
However, with the right hand side depending only on $-v/2M$, we can easily treat Eq.(\ref{f4}) numerically. 
Figure \ref{plot} is a graph of $M$ as a function of $v$.
By the previous study we found that black hole finally vanish at $v=0$ and $M=0$.
The limit of the $-v/2M$ is the solution of the following equation as,
\begin{eqnarray}
(1+e^{-v/2M})+\frac{v}{2M}e^{-v/2M}=0.
\end{eqnarray}
\section{Black hole evaporation without approximations}\label{sec3}
\quad The result of subsection{2-3} use Candela's approximation\cite{C}.
However, we would like to solve the dynamical horizons equation without approximation.
So we use stress-energy tensor of massless scalar field.
By the similar approach, we can obtain following equation as
\begin{eqnarray}
\frac{1}{2}(2M+2Me^{-v/2M})
=\int _0^{r_D}4\pi r_D^2\bigg( \phi_{,ab}-\frac{1}{2}g_{ab}\phi_{,c}\phi^{,c}\bigg)t^al^a dr_D \nonumber \\
+\int _{0}^{M}\frac{5}{2}(1+e^{-v/2M})dM .\label{eq5}
\end{eqnarray}
Here, $\phi$ is the massless scalar field.
The calculations of the integrand of the first term becomes as,
\begin{eqnarray}
\phi_{,ab}t^al^a=-a^{-1}\phi_{,tv}=-a^{-1}\phi_{,vv}
\end{eqnarray}
\begin{eqnarray}
\phi_{,c}\phi^{,c}=\phi_{,t}^2+\phi_{,r^*}^2=2\phi_{,v}^2=2a\phi_{,r}^2.
\end{eqnarray}
Then the first term becomes as
\begin{eqnarray}
\int _0^{r_D}4\pi r_D^2\bigg( \phi_{,ab}-\frac{1}{2}g_{ab}\phi_{,c}\phi^{,c}\bigg)t^al^a dr_D
=\int_0^{r_D}2\pi r_D^2a^{-1}\phi_{,vv}dr_D.
\end{eqnarray}
Then the dynamical horizons equation \ref{eq5} becomes as,
\begin{eqnarray}
\frac{1}{2}(2M+2Me^{-v/2M})
=\int_0^{r_D}2\pi r_D^2a^{-1}\phi_{,vv}dr_D
+\int _{0}^{M}\frac{5}{2}(1+e^{-v/2M})dM .\label{eq6}
\end{eqnarray}
Using the equation (\ref{eq7}) and by derivative of $M$, we can finally obtain following equation as,
\begin{eqnarray}
(1+e^x)\phi_{,xx}+2\phi_{,x}^2=-\frac{-\frac{3}{2}(1+e^{x})-xe^{x}}
{ \pi (1+e^{x})^2e^{-3x}\bigg( 2(1+e^{x})-2xe^{x}\bigg)} . \label{eq8}
\end{eqnarray}
Here, $x=-v/2M$.
This equation is main result of our works in this paper.
This equation is the second order ordinal differential equation with non-linear term.
This equation can not be solved by analytical method, but it can be solved numerical method.
The equation is the simple form of the field equation.

When black hole vanishes, the approximation of the subsection \ref{sec2-3} is correct, then the denominator of the equation (\ref{eq8}) becomes 0.
And the equation (\ref{eq8}) becomes diverge.
It means the derivative of the field diverges, and it means Schwarzschild spacetime becomes Euclidean spacetime.

If the $x$ become infinity, the equation becomes simple as
\begin{eqnarray}
\phi_{,xx}=-\frac{3}{4\pi}.
\end{eqnarray}

Next we consider when black hole vanish.
Using the following relation 
\begin{eqnarray}
1+e^x\to xe^x,
\end{eqnarray}
we can simplify the equation (\ref{eq8}) as
\begin{eqnarray}
xe^x\phi_{,xx}+2\phi_{,x}^2=\frac{5}{4\pi}\frac{e^{3x}}{1+e^x-xe^x}.\label{eq9}
\end{eqnarray}
Because of numerical calculation, we solve 
\begin{eqnarray}
1+e^x-xe^x=0,
\end{eqnarray}
then we obtained the $x=1.27846$.
So the equation (\ref{eq9}) becomes as,
\begin{eqnarray}
a_1\phi_{,xx}+2\phi_{,x}^2=\frac{5}{4\pi}\frac{a_2}{1+e^x-xe^x}.\label{eq10}
\end{eqnarray}
Here, $a_1=4.59108,a_2=46.311$.
More simplicity we put the right hand side of Eq.(\ref{eq10}) as constant $C$, then we can solve this equation.
And the solution becomes
\begin{eqnarray}
\phi (x)\to 2.29554 \log [\cosh (\sqrt{C}\times 0.308035x)].
\end{eqnarray}
For the large $C$ and small $x$ this function behaves like proportional function.
And when black hole banishes, the function $\phi (x)$ becomes $0$. 
\section{Discussions and conclusions}\label{sec4}
\quad We can say black hole finally vanishes by the Hawking effect by our study as we first prove \cite{Sa}.
The reason why we enter freedom of the field is that we would like to check generalized second low of the black hole thermodynamics.
The main result of present paper is Eq.(\ref{eq8}).
By this result, numerical calculation of the field equation can be simplified.
By our study we can show the field closes zero in the limit of the black hole evaporation.
So we can say that the generalized second low of the black hole thermodynamics is collapsed in this semi-classical approach.

The main problem of our work is at the point that our approach is semi-classical one, i.e. the gravity is classical and the field is the quantum one.
So we should quantize gravitation.
And our work may be enlarged to the charged black holes and the Carr black hole.

When the black hole vanishes, the field does not continuous.
So we can say Schwarzcild universe changes Euclidean universe.\\ \\
\begin{center}
{\bf acknowledgments}
\end{center}
We would like to acknowledge A.Hosoya, M.Shiino and S.A.Hayward for comments and discussions. 

\end{document}